\documentstyle[stwol]{article}

\def\be{\begin{equation}}
\def\ee{\end{equation}}
\def\bea{\begin{eqnarray}}
\def\eea{\end{eqnarray}}

\begin{document}

\title{RECENT DEVELOPMENTS IN NON-PERTURBATIVE QUANTUM FIELD THEORY}

\author{ Sergio FERRARA }

\address{Theoretical Physics Division, CERN, 1211 Geneva 23, Switzerland}




\twocolumn[\maketitle\abstracts{
We report on recent advances in the understanding of non-perturbative phenomena
in the quantum theory of fields and strings.}]

\section{Introduction}
In the last couple of years we have been witnessing a series of impressive results
in the framework of two different but related types of quantum theories:
supersymmetric Yang--Mills theories in four dimensions~\cite{aaa} and
superstrings~\cite{bb}.
They have in common space-time supersymmetry.  They also share the property that a
formal perturbative expansion can be defined for these theories, and
``renormalization" can be used in order to extract finite answers from any
perturbative calculation.

These theories enjoy non-renormalization theorems~\cite{cc} depending on the degree
of supersymmetry of the vacuum around which we define the quantum perturbative
series.  

Let us recall the spin content of the light states:
the gauge theory  includes massless quanta with spin 0, 1/2, 1;
superstrings include massless quanta with spin 0, 1/2, 1, 3/2 and 2; in
a suitable limit ($\alpha' \rightarrow 0$, gauge-coupling fixed)
superstrings reproduce ordinary gauge theories.

Exact non-perturbative results of SQCD have been obtained.  Seiberg and
Witten~\cite{dd} suggested an exact expression for the low-energy effective action
for the Coulomb phase of an $N = 2$ SU(2) SYM theory, which may be regarded as an
extension of the Georgi--Glashow SU(2) gauge theory.

Electric--magnetic duality plays a crucial role to solve the theory, to compute
strong coupling phases of this theory (where massless monopoles and dyons
appear), and to prove colour confinement through a magnetic Higgs mechanism with
a monopole condensation (analogue to Meissner effects in superconductors).

The solution of the problem is possible thanks to non-renormalization theorems,
making the complete perturbative computation affordable.

The non-perturbative part, conjecturally due to processes from multi-instanton
transitions, is obtained from a mathematical hypothesis that identifies electric
and magnetic massive states with windings $(n,m)$ of a genus 1 (torus)
elliptic Riemann surface (genus $n$ for an SU($n$) gauge-theory as shown by Klemm,
Lerche, Theisen and Yankielowicz and by Argyres and Faraggi~\cite{ee}).
Two topologically different cycles correspond to electric and magnetic charges.
Quantum massive BPS states $\psi(n,m)$ correspond to distinct topological
configurations $(n,m)$ of the elliptic surface, in particular
$\psi(1,0)=W$-boson, $\psi(0,1) =$ monopole, $\psi(-1,+1)$ = dyon.

\section{Electric--Magnetic Duality and Supersymmetry}

Let us recall the duality of Maxwell equations
\begin{eqnarray}
&\nabla \cdot (E + iB) = \rho_e + i\rho_m = \rho \nonumber\\
&\nabla \wedge (E + iB) - i \frac{\partial}{\partial t} (E + iB) = J_e + iJ_m =
J\nonumber\\
& L = {\rm Re}(E + iB) \cdot (E + iB) = E^2 - B^2~.\nonumber
\end{eqnarray}
Here $E, B$ are the electric and magnetic fields;  $\rho_e(\rho_m), J_e(J_m)$
denote electric (magnetic) charge density and current, respectively. A topological
term $E \cdot B$ may eventually be added to $L$.  The physical observables such as
the energy density $(E + iB) \cdot (E - iB) = E^2 + B^2,$ and the momentum density
$(E + iB) \wedge (E - iB)$ are invariant under (continuous) U(1) duality rotations:
\[
(E + iB) \rightarrow e^{i\varphi}(E + iB)~, ~~ \rho \rightarrow
e^{i\varphi}\rho~,~~J \rightarrow e^{i\varphi}J~.
\]
In particular the $Z_2$ symmetry, which is the remnant of U(1), acting on discrete
charged states, exchanges electric with magnetic fields $E \rightarrow B~{\rm
and}~B \rightarrow -E,$ and electric and magnetic charges $q
\rightarrow g, g \rightarrow -q$, accordingly.

The simultaneous occurrence of electric and magnetic sources implies a charge
quantization, which reads:
\begin{eqnarray}
&qg = 2\pi k\hfill \nonumber \\
&{\rm\hbox{ (Dirac~1931)~\cite{ff}}} {\rm (for~monopoles)}\nonumber
\end{eqnarray}
and
\begin{eqnarray}
&q_1g_2 - q_2g_1 = 2\pi k \nonumber \\
&{\rm
\hbox{(Schwinger,~Zwanziger~1968)~\cite{ggg}}}
 {\rm (for~dyons).}\nonumber
\end{eqnarray}
In the Coulomb phase the Georgi--Glashow SU(2) gauge theory has a monopole with
mass ('t~Hooft, Polyakov 1974)~\cite{hh}:
\begin{eqnarray}
&M_{{\rm monopole}} \geq 1/\lambda \langle \phi \rangle\nonumber \\
&{\rm\hbox{
(Bogomolny~bound, ~1975)~\cite{jj}}}\nonumber
\end{eqnarray}
while the classical vector boson mass is $M_W = \lambda \langle \phi \rangle$.
In the Prasad--Sommerfeld (1976) limit~\cite{kk} (supersymmetry)
$M_{\rm monopole} = 1/\lambda \langle \phi \rangle$
satisfies the duality conjecture (Montonen, Olive 1977)~\cite{lll}:
\[
M^2(q,g) = M^2(q^2 + g^2) = \langle \phi \rangle^2 (q^2 + g^2)~.
\]
This generalizes when a topological term $\theta E\cdot B$ is included by defining
a complex parameter $\tau = \theta + i/\lambda^2$ and then writing:
\[
M^2(\phi,\tau,n,m) = \frac{|\phi^2|}{lm\tau} |n + \tau m|^2
\]
invariant under $SL(2,Z)$:
\begin{eqnarray}
Z_2 &:& \tau \rightarrow \frac{-1}{\tau}~,~~n \rightarrow m~, ~~m \rightarrow -n
\nonumber\\
\theta-{\rm shift}&:& \tau \rightarrow \tau +1~,~~n\rightarrow n-m~,~~ m
\rightarrow m~.\nonumber
\end{eqnarray}
This means that the dual theory obtained by a $Z_2$ symmetry $E \rightarrow B, B
\rightarrow -E$ has $\tau_d = -1/\tau, n_D = m, m_D = -n$.
The $N=4$ supersymmetric Yang--Mills theory realizes the Montonen--Olive duality
conjecture~\cite{lll}.  The theory has an exact
SL(2,$Z$) symmetry~\cite{dda}$^,$\cite{dd}, which is possible in virtue of a
vanishing
$\beta$ function, in the full quantum theory.  
Electric states are fundamental, while magnetic states are solitons in the theory
$T$, but their role is reversed in the dual theory $T_D$.

Seiberg and Witten~\cite{dd} extended the duality to $N=2$, SYM quantum field
theories undergoing renormalization $(\beta \not= 0)$, which gives corrections to
a `holomorphic prepotential', $F(\phi)$;  this is the appropriate tool to build
up
$N=2$ effective actions.  The BPS states (which lie in hypermultiplets) have mass
$M(\phi,n,m,\lambda) \propto |\phi n +F_\phi m|$ where~\cite{mm}
$F(\phi) = (i/2\pi)\phi^2 \ln(\phi^2/\lambda^2) + \dots$ (the dots denote the
non-perturbative contributions).

They also extended the duality conjecture.  This came by identifying the pair
($\phi, F_\phi)$ with the periods of an hyper-elliptic surface, which allows us
to give a closed expansion for $F(\phi)$.  As a result of this at strong coupling
$\phi^2/\lambda^2 =
\pm 1$, one gets a massless monopole (0,1) and a dyon ($-1,+1$).

The dual (U(1) magnetic) theory is weakly coupled in the strong coupling of the
electric theory and describes a magneto-dynamic of a charged monopole.  In the
weakly coupled magnetic Higgs phase, monopole condensation describes confinement of
the original (strongly coupled) electric theory.  It is worth mentioning that, for
BPS states,  their mass appears in the central extension of the supersymmetry
algebra (Haag--Lopuszanski--Sohnius)~\cite{nn} and this allows one, using
supersymmetry~\cite{oo}, to compute their mass in terms of the low-energy data. 
The duality has been further extended to $N=1$ super-Yang--Mills
theories~\cite{pp}, in particular to SQCD with colour group SU$(N_c)$ and $N_f$
flavours.  This theory has an anomaly-free global symmetry:
\[
SU_L(N_f) \times SU_R(N_f) \times U(1)_B \times U(1)_R~.
\]
Seiberg suggested that there is a non-Abelian Coulomb phase for $3/2N_c < N_f <
3N_c$.  At the non-trivial infra-red fixed point, the theory of quarks and gluons
has a dual description in terms of an interacting conformal invariant theory with
magnetic gauge group $SU(N_f-N_c)$ and $N_f$ flavours.
Quarks and gluons are solitons in the dual picture.

\section{Supergravity, Strings and $M$-Theory}

Duality symmetries in the context of supergravity theories~\cite{qq}, further
extended to superstrings~\cite{rr}, allow us to prove exact equivalences of different
string theories~\cite{ss}$^{\rm\hbox{--}}$\cite{uu}, to obtain a dynamical
understanding of the Seiberg--Witten conjecture in the point-particles
limit~\cite{vv}$^{\rm\hbox{--}}$\cite{ww} and finally to possibly merge these
theories in the context of $M$-theory, a supposedly existing quantum theory of
membranes and five-branes~\cite{yy}$^,$\cite{zz}, whose low-energy effective action is
11D supergravity~\cite{aai}.

There are five known types of superstring theories in 10 dimensions~\cite{bb}:

\begin{tabular}{l | l }
Type & Gauge group\\ \hline
Type I & SO(32)\\ 
Heterotic & SO(32), $E_8 \times E_8$\\
Type IIA & U(1)\\
Type IIB & None
\end{tabular}

The first three have $N = 1$ supersymmetry, while the last two have $N = 2$,
non-chiral type IIA and chiral IIB. There is also a conjectured 
$M$-theory in 11 dimensions~\cite{uu} (no gauge group).  Upon reduction on a circle,
this is equivalent to type IIA, at the non-perturbative level.  A further
speculative theory may exist in twelve dimensions, which gives, upon reduction on a
two-torus, the type IIB theory~\cite{bbi}.

The previous theories, and their compactification to lower dimensions, reduce at
low energies to supergravity theories in diverse dimensions~\cite{cci} with 
underlying supersymmetry algebras as classified by Nahm~\cite{ddi}.  In the
highest and lowest dimensions of interest we have for instance:
\begin{eqnarray}
D =   11, N &= &1, 128_{\rm boson} + 128_{\rm fermions}\nonumber\\
&&(b = 44 + 84, f =
128)\nonumber \\
D = 10, N & = &1~({\rm chiral})\nonumber \\
N &=& 1~({\rm matter}) (G = E_8 \times E_8, {\rm SO(32))}\nonumber \\
N &= & 2: {\rm\hbox{ type~A~(non-chiral)}},\nonumber \\
&& {\rm\hbox{type~B~(chiral)}}\nonumber\\
 D=4,  N&=&1~{\rm (chiral):~obtained~as}~M{\rm\hbox{-theory}}\nonumber\\
&&{\rm on}~M_7 = CY_3
\times S_1/Z_2 \nonumber\\
N&=&8~{\rm\hbox{(non-chiral)}}~(b = 56 + 70 + 2,\nonumber\\
&&f = 112 + 16):~{\rm obtained~as}~\nonumber \\
&& M{\rm\hbox{-theory on}}~T_7({\rm U}(1)^{28}~{\rm gauge}\nonumber\\
&& {\rm group)}\nonumber \\
&&{\rm or~on}~S_7({\rm SO}(8)~{\rm gauge~group)}~.\nonumber
\end{eqnarray}

Let us summarize some of the main
basic results of the years 94--96, in the context of string theory and
its non-perturbative regime.

\begin{itemize}
\item[1)] The Seiberg--Witten solution of rigid $N=2$ theory generalizes
to heterotic-type II duality relating $K_3 \times T_2$ vacua of heterotic
to Calabi--Yau vacua of type II strings~\cite{eei}$^,$\cite{ffi}. 

The second quantized mirror symmetry~\cite{eei} gives exact
non-perturbative results in
$N=2$ superstrings, $D = 4$.  In particular, duality relates world-sheet
instanton effects on the type II side to space-time instantons on the
heterotic side~\cite{vv}$^,$\cite{ggi}. 
Dual pair heterotic-type II theory constructions were proposed.

\item[2)] The implication of string--string duality in six dimensions for $S$-$T$
duality at $D = 4$ was first shown by Duff~\cite{ss}, and
$U$-duality as a non-perturbative symmetry of different string theories
was formulated by Hull and Townsend~\cite{tt}.

\item[3)] Witten~\cite{uu} proved the equivalence of different string
theories in higher dimension and the duality of type IIA at strong
coupling with 11$D$ supergravity at large radius ($M$-theory on
$M_{10}\times S_1$).  Type IIB is self-dual at $D = 10$ (SL$(2,Z)$
duality)~\cite{hhia}.
 
\item[4)] The $E_8 \times E_8$ heterotic string at strong coupling is dual
to the $M$-theory on $M_{10} \times S_1/Z_2$ (Horava--Witten)~\cite{hhi}.

\item[5)] The SO(32) Type I and SO(32) heterotic at $D =
10$ are interchanged by  weak--strong coupling duality
(Polchinski--Witten)~\cite{jji}.

\item[6)] Open strings naturally arise, by the mechanism of tadpole
cancellations, as sectors of type IIB closed strings on
orientifolds~\cite{kki}$^,$\cite{lli}.  Their end-points end on
$D$-branes~\cite{mmi}, carrying R--R charges.  Phase transitions in six
dimensions are possible~\cite{nni}, and evidence for a non-perturbative
origin of gauge symmetries~\cite{ooi} was substantiated~\cite{ppi}.

\item[7)] $T$-duality at $D = 9$ relates type IIA and type IIB
theories, as well as SO(32) and $E_8 \times E_8$ heterotic strings in
their broken phase SO(16)$\times$SO(16)~\cite{jji}.

\item[8)] $M$-theory and strings may undergo a further
unification in twelve dimensions ($F$-theory)~\cite{bbi}.
\end{itemize}

New predictions of non-perturbative string\break theories can be derived from
these non-\break perturbative relations between the five seemingly different
superstring theories.  As a circumstantial example \cite{qqi}, 
strongly coupled heterotic string meets the agreement of $\alpha_{GUT}$ as
measured (at LEP) from low-energy data.

In weakly coupled heterotic string, compactified on a Calabi--Yau threefold of size
$V \approx M_{GUT}^{-6}$ with $G_N = \frac{e^{2\phi}(\alpha')}{64\pi V}$~,
\[
~\alpha_{GUT} = e^{2\phi}(\alpha')^3/16\pi V \rightarrow G_N =
\alpha_{GUT}\alpha'/4~.
\]
If $e^{2\phi} \leq 1, G_N \geq \alpha_{GUT}^{4/3}/M^2_{GUT}$, which is too large
compared to experiment.

In type I string (weak coupling)~,
\begin{eqnarray}
\alpha_{GUT} &= &\frac{e^{\phi_I}(\alpha')^3}{16\pi V},~G_N =
\frac{e^{2\phi_I}(\alpha')}{64 \pi V}\nonumber \\
 &\rightarrow& G_N = e^{\phi_I}\alpha_{GUT}
\alpha'/4~.\nonumber
\end{eqnarray}
Here, $G_N$ can be small.

  In the $M$-theory set-up ($\kappa$ is the 11$D$ gravitational coupling and $\rho$
the compactification radius)
\[
G_N = \frac{k^2}{16\pi^2V\rho}~,~~\alpha_{GUT} = \frac{(4\pi k^2)^{2/3}}{2V}\ll
1~.
\]
So no disagreement with the experimental input exists in principle.

Finally, supersymmetry breaking can be described in a natural way both through a
strongly coupled hidden gauge sector leading to gaugino condensation~\cite{rri} and
through the no-scale structure~\cite{ssi} of $M$-theory.  The decompactification
problem may be avoided~\cite{tti}$^{\rm\hbox{--}}$\cite{vvi}.

\section{Conclusions}

In the talk he gave at the Stony-Brook conference in 1979, Murray
Gell-Mann~\cite{wwi} outlined the problems encountered with
$N = 8$ supergravity~\cite{wwia}:
\begin{itemize}
\item[1)]
``Predict a gauge group SO(8) that does not contain SU(3)$\times$ SU(2) $\times$
U(1);
\item[2)] `Sign error' in the prediction of the cosmological constant
$\lambda \approx M^4_{Planck}$!
\item[3)] We can use lower-$N$ supersymmetric theories, but then we have a lack of
unification~..."
\end{itemize}

It may be surprising to realize that the unphysical theory discussed by Gell-Mann
17 years ago is just a different vacuum of the very same theory whose dynamics
hopefully encodes the standard model of our low-energy world!

\section*{Questions}
\noindent{\it G.G. Ross, Oxford University:}

What progress has there been towards breaking supersymmetry and using duality to
give information about normal QCD?

\vskip 12pt
\noindent {\it S. Ferrara:}

Alvarez-Gaum\'e, Distler, Kounnas and Mari$\tilde{\rm n}$o~\cite{xxi} have analysed
soft breaking terms preserving the analyticity properties of the Seiberg--Witten
solution.  This allows a detailed description of the onset of the confinement
transition and the pattern of chiral symmetry breaking.  When those results are
extrapolated to a limit where supersymmetry decouples and then QCD is retrieved, an
indication that the QCD vacuum may require the simultaneous occurrence of mutually
non-local degrees of freedom (monopoles and dyons) seems to emerge.

\vskip 12pt
\noindent {\it G. Veneziano, CERN:}
In order to get a successful relation between the Planck scale and the GUT scale
is it crucial that the original type II theory is strongly coupled or can the
coupling be just at the self-dual value = 1?

\vskip 12pt
\noindent {\it S. Ferrara:}

The strong coupling in heterotic theory means that its dual theory is weakly
coupled.  Therefore the self-dual value seems not appropriate for this regime.


\section*{References}
\begin{thebibliography}{99}
\bibitem{aaa} For reviews on Supersymmetric Gauge Theories, see for instance:\\
J. Bagger and J. Wess, {\it Supersymmetry and Supergravity} (Princeton University
Press, 1991);\\
S. Ferrara, {\it Supersymmetry} (North Holland--World Scientific, 1987).
\bibitem{bb} For a review on Superstrings, see:  M. Green, J. Schwarz and
E. Witten, {\it Superstring Theory} (Cambridge University Press, 1987).
\bibitem{cc} J. Iliopoulos and B. Zumino, {\it Nucl. Phys.} {\bf B76} (1974)
310;\\
S. Ferrara, J. Iliopoulos and B. Zumino, {\it Nucl. Phys.} {\bf B77} (1974) 413;\\
M.T. Grisaru, W. Siegel and M. Ro\v cek, {\it Nucl. Phys.} {\bf B159} (1979)
429;\\
L. Girardello and M.T. Grisaru, {\it Nucl. Phys.} {\bf B194} (1982)
65.
\bibitem{dd} N. Seiberg and E. Witten, {\it Nucl. Phys.} {\bf B426} (1994) 19 and
{\bf B431} (1994) 484.
\bibitem{ee} A. Klemm, W. Lerche, S. Theisen and S. Yankielowicz, {\it Phys.
Lett.} {\bf B344} (1995) 169;\\
P. Argyres and A. Faraggi, {\it Phys. Rev. Lett.} {\bf 74} (1995) 3931.
\bibitem{ff} P.A.M. Dirac, {\it Proc. Roy. Soc.} {\bf A33} (1931) 60.
\bibitem{ggg} J. Schwinger, {\it Science} {\bf 165} (1969) 757;\\
D. Zwanziger, {\it Phys. Rev.} {\bf 176} (1968) 1489.
\bibitem{hh} G. 't Hooft, {\it Nucl. Phys.} {\bf B79} (1974) 276;\\
A.M. Polyakov, {\it JETP. Lett.} {\bf 20} (1974) 194.
\bibitem{jj} E.B. Bogomolny, {\it Sov. J. Nucl. Phys.} {\bf 24} (1976) 449.
\bibitem{kk} M.K. Prasad and C.M. Sommerfeld, {\it Phys. Rev. Lett.} {\bf 35}
(1975) 760.
\bibitem{lll} C. Montonen and D. Olive, {\it Phys. Lett.} {\bf 72B} (1977) 117.
\bibitem{dda} L. Girardello, A. Giveon, M. Porrati and A. Zaffaroni, {\it Phys.
Lett.} {\bf B334} (1994) 331.
\bibitem{mm} N. Seiberg, {\it Phys. Lett.} {\bf B206} (1988) 75.
\bibitem{nn} R. Haag, J.T. Lopuszanski and M. Sohnius, {\it Nucl. Phys.} {\bf
B88} (1975) 257.
\bibitem{oo} E. Witten and D. Olive, {\it Phys. Lett.} {\bf 78B} (1978) 97.
\bibitem{pp} N. Seiberg, {\it Phys. Rev.} {\bf D49} (1994) 6857.
\bibitem{qq} E. Cremmer, S. Ferrara and J. Scherk, {\it Phys. Lett.} {\bf 74B}
(1978) 61;\\
S. Ferrara, J. Scherk and B. Zumino, {\it Nucl. Phys.} {\bf B121} (1977) 393.
\bibitem{rr} A. Sen, {\it Int. J. Mod. Phys.} {\bf A9} (1994) 3707;\\
J.H. Schwarz, {\it Lett. Math. Phys.} {\bf 34} (1995) 309.
\bibitem{ss} M. Duff, {\it Nucl. Phys.} {\bf B442} (1995) 47.
\bibitem{tt} C. Hull and P. Townsend, {\it Nucl. Phys.} {\bf B438} (1995) 109.
\bibitem{uu} E. Witten, {\it Nucl. Phys.} {\bf B443} (1995) 85.
\bibitem{vv} A. Ceresole, R. D'Auria and S. Ferrara, {\it Phys. Lett.} {\bf 339B}
(1994) 71;\\
A. Ceresole, R. D'auria, S. Ferrara and A. van Proeyen, {\it Nucl. Phys.} {\bf
B444} (1995) 92.
\bibitem{ww} S. Kachru, A. Klemm, W. Lerche, P. Mayr and C. Vafa, {\it Nucl.
Phys.} {\bf B459} (1996) 537.
\bibitem{yy} M.J. Duff and J.X. Lu, {\it Nucl. Phys.} {\bf B426} (1994) 301.
\bibitem{zz} E. Witten, preprint IASSNS-HEP-9563 (hep-th/9507121), contribution
to Strings '95, Los Angeles, 1995.
\bibitem{aai} E. Cremmer, B. Julia and J. Scherk, {\it Phys. Lett.} {\bf 76B}
(1978) 409.
\bibitem{bbi} C. Vafa, {\it Nucl. Phys.} {\bf B469} (1996) 403.
\bibitem{cci} A. Salam and E. Sezgin, {\it Supergravity in diverse dimensions}
(North-Holland--World Scientific, 1989).
\bibitem{ddi}W. Nahm, {\it Nucl. Phys.} {\bf B135} (178)
149.
\bibitem{eei} S. Ferrara, J. Harvey, A. Strominger and C. Vafa, {\it Phys. Lett.}
{\bf B361} (1995) 59.
\bibitem{ffi} S. Kachru and C. Vafa, {\it Nucl. Phys.} {\bf B450} (1995)
69.
\bibitem{ggi} A. Strominger, {\it Nucl. Phys.} {\bf B451} (1995) 97;\\
B. Greene, D. Morrison and A. Strominger, {\it Nucl. Phys.} {\bf 451} (1995) 109.
\bibitem{hhi} P. Horava and E. Witten, {\it Nucl. Phys.} {\bf B460} (1996) 506.
\bibitem{hhia} For recent reviews. see:\\
J. Schwarz, preprint CALTECH-68-2065 (hep-th/9607201);\\
A. Sen, preprint MRI-PHY-96-28 (hep-th/9609176);\\
M. Duff, preprint CPT-TAMU-33-96 (hep-th/9608117).
\bibitem{jji} J. Polchinski and E. Witten, {\it Nucl. Phys.} {\bf B460} (1996)
525.
\bibitem{kki} A. Sagnotti, {\it Open Strings and their Symmetry Groups}, talk at
the Carg\`ese Summer Institute (1987);\\
G. Pradisi and A. Sagnotti, {\it Phys. Lett.} {\bf B216} (1989) 59;\\
M. Bianchi, G. Pradisi and A. Sagnotti, {\it Nucl. Phys.} {\bf B376} (1992) 365.
\bibitem{lli} J. Dai, R. Leigh and J. Polchinski, {\it Mod. Phys. Lett.} {\bf A4}
(1989) 2073;\\
P. Horava, {\it Nucl. Phys.} {\bf B227} (1989) 461;\\
J. Polchinski, {\it Phys. Rev.} {\bf D50} (1994) 6041.
\bibitem{mmi} J. Polchinski, {\it Phys. Rev. Lett.} {\bf 75} (1995) 4734.
\bibitem{nni} A. Sagnotti, {\it Phys. Lett.} {\bf 294B} (1992) 196.
\bibitem{ooi} E. Witten, {\it Nucl. Phys.} {\bf B471} (1996) 121.
\bibitem{ppi} M. Duff, R. Minasian and E. Witten, {\it Nucl. Phys.} {\bf B465}
(1996) 44.
\bibitem{qqi} E. Witten, {\it Nucl. Phys.} {\bf B471} (1996) 121.
\bibitem{rri}P. Horava, preprint PUPT-1637 (hep-th/9608019).
\bibitem{ssi} E. Cremmer, S. Ferrara, C. Kounnas and D.V. Nanopoulos, {\it Phys.
Lett.} {\bf B133} (1983) 61;\\
J. Ellis, A. Lahanas, D.V. Nanopoulos and K. Tamvakis, {\it Phys. Lett.} {\bf
B134} (1984) 419;\\
J. Ellis, C. Kounnas and D.V. Nanopoulos, {\bf B241} (1984) 406.
\bibitem{tti} T. Banks and M. Dine, preprint SCIPP-96-26 (hep-th/9609046); 
preprint RU-96-27 (hep-th/9605136).
\bibitem{uui}E. Kiritsis, C. Kounnas, P.M. Petropoulos and J. Rizos, preprint
CERN-TH/96-90 (hep-th/9608034).
\bibitem{vvi} I. Antoniadis and M. Quiros, CPTH-S465-0996 (hep-th/9609209).
\bibitem{wwi} M. Gell-Mann in {\it Supergravity}, eds. D.Z. Freedman and P. van
Nieuwenhuizen (North-Holland, Amsterdam, 1979), p. 315.
\bibitem{wwia} J. Cremmer and B. Julia, {\it Nucl. Phys.} {\bf B159} (1979) 141;\\
B. de Wit and H. Nicolai, {\it Nucl. Phys.} {\bf B208} (1982) 323.
\bibitem{xxi} L. Alvarez-Gaum\'e, J. Distler, C. Kounnas and M. Mari$\tilde{\rm
n}$o, {\it Int. J. Mod. Phys.} {\bf A11} (1996) 4745.
\end{thebibliography}
\end{document}